\documentclass[prl,twocolumn,showpacs]{revtex4}
\usepackage{graphicx}
\usepackage{bm}
\usepackage{epsfig}

\begin{document}

\title{Impact of dissipative effects on the 
macroscopic evolution of a Vlasov system}

\author{L. Galeotti, F. Califano}
\affiliation{Physics Dept., University of Pisa and INFM, Pisa, Italy}
\date{\today}
\begin{abstract}
Numerical diffusion is introduced by any numerical scheme 
as soon as small scales fluctuations, generated during the 
dynamical evolution of a collisionless plasma, become comparable 
to the grid size. Here we investigate the role of numerical 
dissipation in collisionless plasma simulations by studying 
the non linear regime of the two  stream instability. 
We show that the long time evolution of the Vlasov - Poisson 
system can be affected by the used algorithm. 
\end{abstract}

\pacs {52.65.Ff, 52.35.Qz, 52.35.Fp, 52.35.Mw}

\maketitle

\noindent
Many space and laboratory plasmas can be considered as
weakly collisional since the collisional frequency is
smaller than all the other frequencies, as
for example the plasma frequency. In other words,
for these plasmas, the mean free path of the 
particles is (much) longer than all the other
characteristic length scales of the plasma and, sometimes, even
larger than the dimension of the plasma itself. At first
approximations such plasmas can be considered as collisionless and
their dynamics can be well represented using a Hamiltonian
description. This approach is based on the idea that the
dissipative scale (for example in numerical simulations the grid
size) is much smaller than any macroscopic physical length scale
of the system, so that dissipation has no feedback on the macroscopic
asymptotic evolution of the system. 

Numerical simulations
based on non collisional models must necessarily face 
with the small scales generation problem during
the dynamical evolution of the system; indeed, when
the typical length scales of the fluctuations become
comparable to the grid size, numerical dissipation comes into play
leading the system to violate the conservation constraints of
Hamiltonian dynamics and to reconnect close isolines of the
distribution function (d.f.). This process, formally forbidden, is
very well highlighted by the time evolutions of the system
invariants $N_i = \int f^i dx dv \,\,\,\,  i = 1, 2,..$ and by the
"entropy" ${\cal S} = - \int f \, ln(f) dx dv$ (here $f$ is the
d.f.), showing sudden variations when closed
vortices are formed in phase space as a consequence of particle 
trapping. 

In this paper, through
numerical studies of a non-linear regime of a collisionless
plasma, we discuss the role of artificial dissipation
introduced by a numerical scheme on the plasma dynamics,
influencing the final Vlasov evolution of the
system even if the grid size is much
shorter than any physical relevant scale length. The dynamical 
non linear evolution we chose for our numerical simulations is
the well-known two stream instability. In this case, 
dissipation allow for the formation of coherent 
macroscopic structures in phase space (vortices).

Since the non-linear dynamics of the two stream instability is
substantially driven by kinetic effects, especially concerning the
saturation phase where particle trapping play a crucial role, a
kinetic approach is necessary. This can be done using Vlasov
equation, which replaces Coulomb interactions between charged
particles with a mean electromagnetic field. This field is
determined self-consistently trough the particle distribution
function by Maxwell and Poisson equations. Since the two stream
instability is driven by purely electrostatic mechanisms, we limit
our study to the solution of the 1D-1V Vlasov - Poisson system of
equations:

\begin{equation}  \label{eq:Vlasov}
\frac{\partial f_a}{\partial t} + v\, \frac{\partial f_a}{\partial
x} - \frac{m_e}{m_a} \, \frac{\partial \phi}{\partial x}\,
\frac{\partial f_a}{\partial v} = 0; \;\;\;\; a = e,\, p \,\,
\end{equation}
\begin{equation}  \label{eq:poisson}
\frac{\partial^2 \phi}{\partial x^2} =  \int \left( f_e - f_p
\right) \, dv; \;\;\;\; E = - \frac{\partial \phi}{\partial x}
\end{equation}

In these equations and in the rest of this paper, time $t $
is normalized to the inverse of the electron plasma frequency 
$\omega_{pe}$, velocities $v_e$ and $v_p$ to the electronic
thermal velocity $v_{th,e}$, electron and proton distribution
functions, $f_e$ and $f_p$, to the equilibrium particle
density $n_0$, lengths to the Debye length
$\lambda _D = v_{th,e}/\omega _{pe}$ and the electric field
$E$ to $m_e v_{th,e} \omega _{pe} /e$. As mentioned above, the
dynamics described by this system of equations is characterized by
the absence of collisions; hence, from a numerical point of view,
the choice of an algorithm that is capable to conserve better than
possible the invariants of the system is crucial. Our numerical
scheme is based on the splitting scheme of Cheng and Knorr, 1976,
\cite{Cheng} for the solution of the Vlasov equation; therefore,
the problem is mainly reduced to an interpolation
problem for the distribution function. Here we compare
three well known interpolation algorithms, namely
the Van Leer method at second and third order \cite{Mangeney}
(at which, in the text, we'll refer as VL2 and VL3) and
the Spline method \cite{Shoucri} (a third order method).
The Poisson equation is solved, at every time step, by
spectral methods (i.e. fast Fourier Transform technique);
in particular, we calculate the plasma density by integrating
the electron distribution functions in velocity.

\begin{figure*}[!t]
\centerline{\psfig{figure=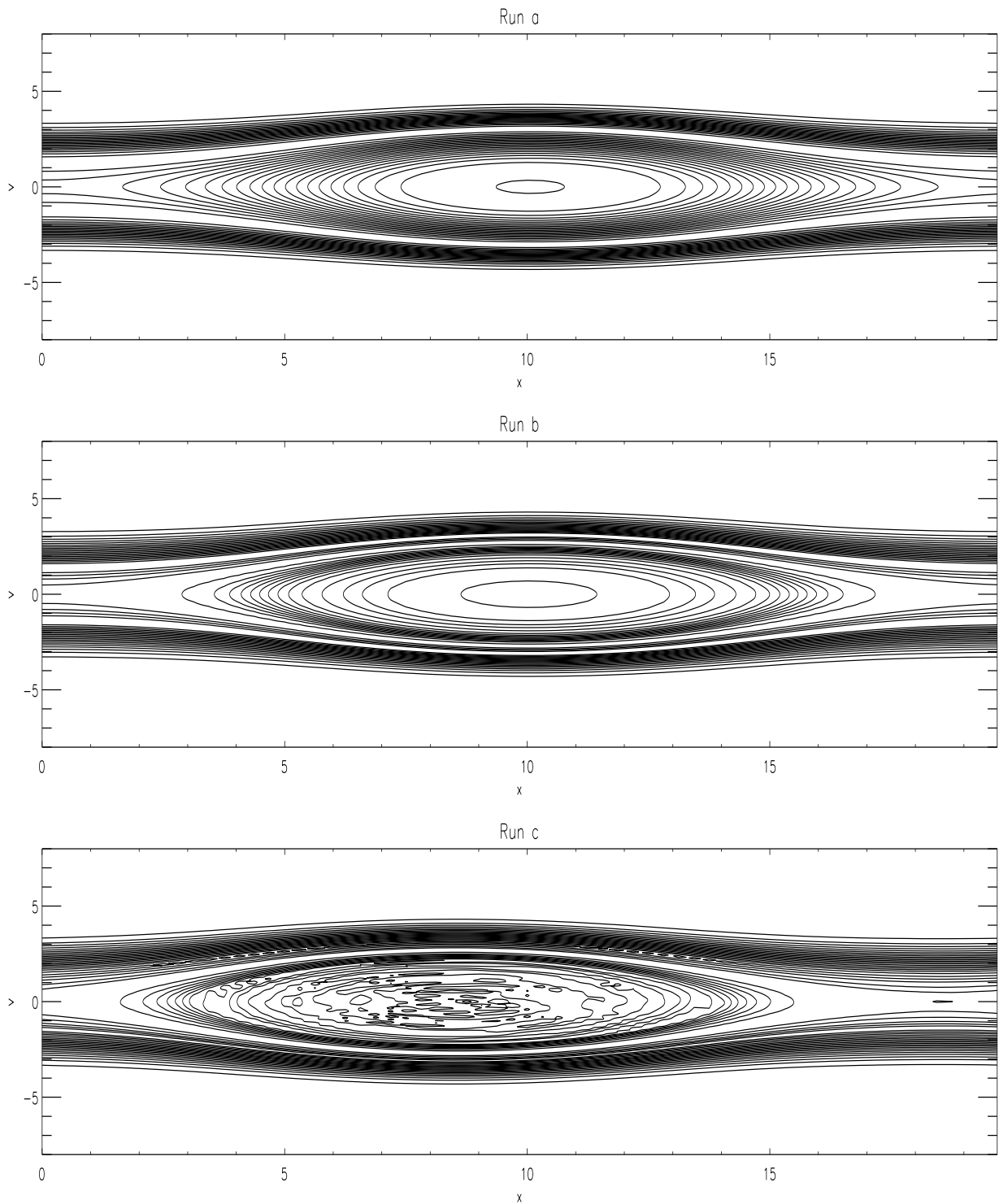,height=20cm,width=17cm}}
\caption[ ]{Electron distribution functions in phase space at time
t=1200 of, top to bottom, the VL2, VL3 and Spline algorithm,
runs A, B and C, respectively. }
\label{fig1}
\end{figure*}

We made the two stream instability runs for the three simulation
algorithms (VL2, VL3 and Spline, runs A, B and C, respectively)
using the same identical
parameters (however, we recall that the CPU time for the three
algorithms is not equal). The simulations box is
$L_x = 20 \lambda_D$ in space and the velocity interval is
$- v_{max} \leq v \leq v_{max}$, with $v_{max} = 5 \, v_{th,e}$.
We use $N_x=128$ points in space and $N_v=501$ in velocity,
corresponding to a phase space grid resolution of
$dx = 0.16$ and $dv = 0.02$.
Other parameters are: amplitude of the initial random perturbation
of $\epsilon = 0.0001$; $dt=0.0003 \, \omega _{pe}^{-1}$, total
simulation time equal to $1200 \, \omega _{pe}^{-1}$; modulo of mean
velocity of the two initial electron streams ($u_0$) equal to 2 $v_{th,e}$.
The initial electron distribution function we use is
\begin{equation}\label{Vlas1}
 f_e \left(x,v_e,0 \right) = f_M \left(v_e \right)
\left[ 1 + \epsilon
\sum_{k=1}^{30} cos \left(kx + \phi _k \right)  \right]  \\
\end{equation}
\begin{equation}\label{Vlas2}
f_M = \frac{1} {\sqrt{2\pi} \, v_{th}} \, (e^{-( (v_e-v_0)^2
/ 2 v_{th}^2 )}+ e^{-( (v_e+v_0)^2 / 2 v_{th}^2 ) })
\end{equation}
In Fig. \ref{fig1} we show the phase space vortices (same contour
levels) generated by the evolution of the instability.
\begin{figure}[!t]
\centerline{\psfig{figure=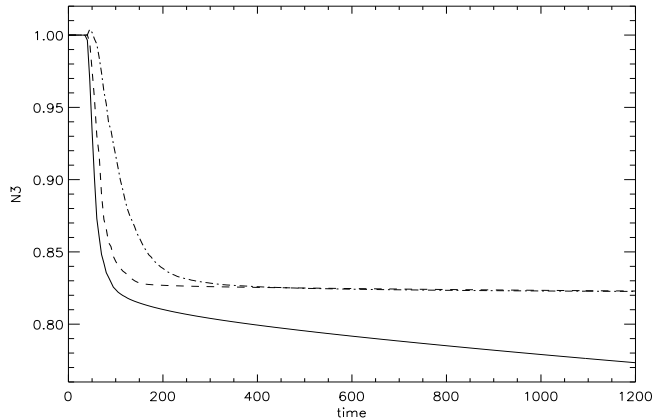,height=6cm,width=9cm}}
\caption[]{The time evolution of the third invariant. The continuous,
dashed and dot-dashed lines correspond to run A to C,
respectively. }
\label{fig2}
\end{figure}
There is an initial good agreement among the results of the three
algorithms: the vortex appears at the same time for all methods
and is displaced in the same position. On the other hand, we
observe that the Spline vortex (run C) propagates with a different
velocity with respect to the vortex obtained with VL2 and VL3
(runs A, B) method. Furthermore, the spatial structure of
the VL2 vortex is significantly different from that of
the vortices obtained with the other two algorithms.

\begin{figure}[!t]
\centerline{\psfig{figure=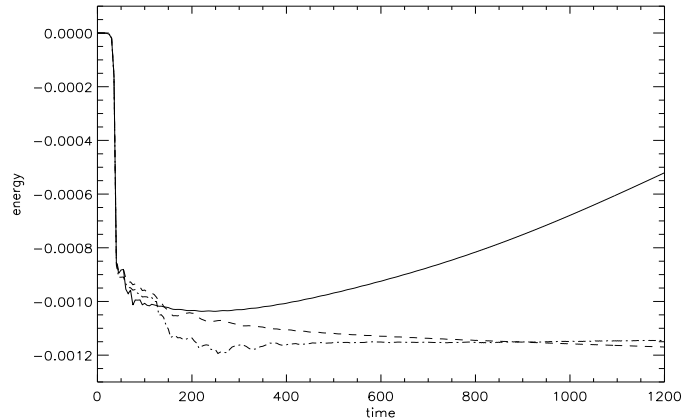,height=6cm,width=9cm}}
\caption[ ]{The time evolution of $\delta E$ (same line style as
in fig. \ref{fig1}. }
\label{fig3}
\end{figure}

\begin{figure}[!t]
\centerline{\psfig{figure=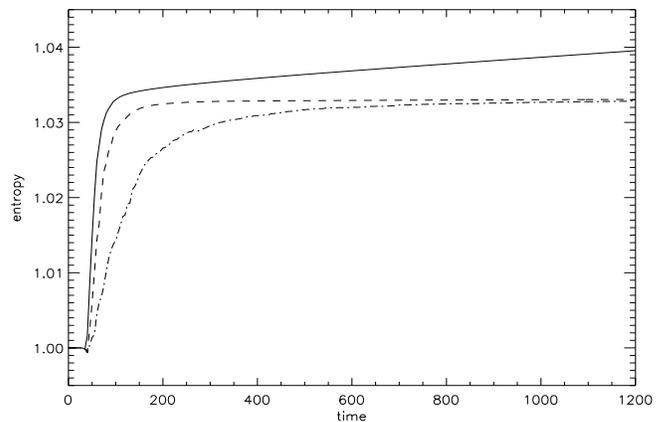,height=6cm,width=9cm}}
\caption[ ]{The time evolution of the entropy (same line style as
in fig. \ref{fig2}. }
\label{fig4}
\end{figure}

By looking at the behavior of the $N_3$ invariant, Fig.
\ref{fig2}, we see in all cases a sudden decrease of the invariant
as soon as the vortices start to form. This is a consequence of
the d.f. lines reconnection processes at the grid scale length
where the algorithm is forced to introduce some artificial
dissipation eventually leading to the closure of the particle
orbits (i.e. vortices) contrary to the Hamiltonian character of
the equations. We note that both VL3 and Spline invariants tend to
a (equal) constant asymptotic value, while VL2 continues to
smoothly decrease, meaning that the numerical resolution for the
VL2 algorithm must be increased (even if both $dx$ and $dv$ are
much shorter than the vortex dimension), as also shown by the time
evolution of the total energy variations of the system in Fig.
\ref{fig3} where we plot the evolution of the normalized energy
variations defined as $\delta E = (E_{tot} - E_{tot}(t=0)) /
E_{tot}(t=0)$. Indeed, we see that after the phase space vortex
formation, the energy variation for VL3 and Spline algorithms
becomes nearly constant, while for VL2 begins to monotonically
increase. In Fig. \ref{fig4} we show the time evolution of the
entropy. We again observe a strong variation of entropy during the
vortex formation phase for all three algorithms while in the
asymptotic limit the entropy becomes nearly constant for both VL3 and
Spline, while continues to increase for VL2.
\begin{figure}[!t]
\centerline{\psfig{figure=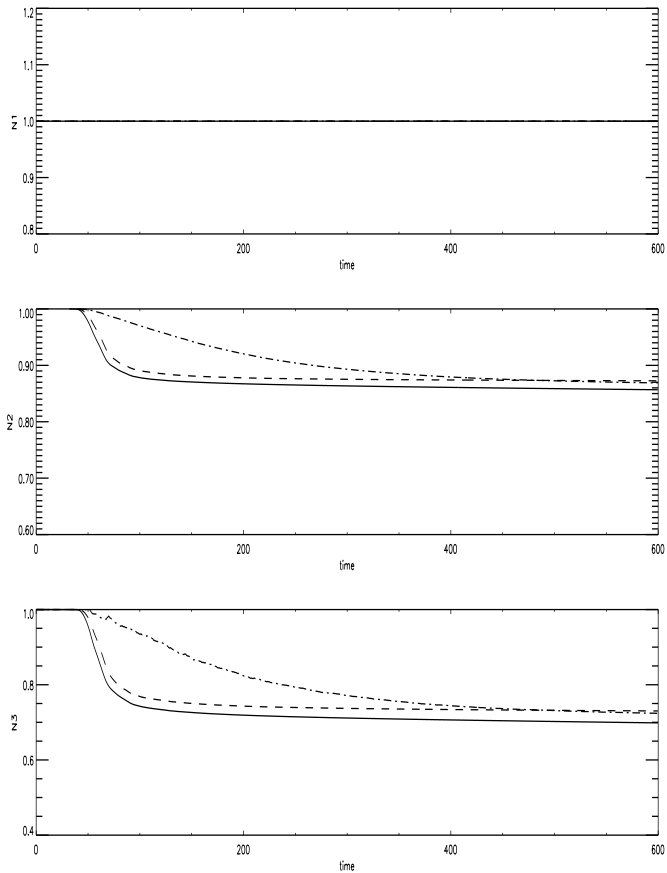,height=12cm,width=9cm}}
\caption[ ]{The time evolution of the invariants $N_1$ 
(total charge density), $N_2$ and $N_3$. The
continuous, dashed and dot-dashed lines correspond to runs
D, E and F made with the VL2, VL3 and Spline algorithm,
respectively. }
\label{fig5}
\end{figure}
\noindent
The uncorrect behavior of the VL2 method with respect to VL3 and
Spline is the consequence of the fact that VL2 method is a lower
order scheme. We underline that VL2 is however a II order scheme
(II order schemes are often used in collisionless simulations) and
that the grid spacing, $dx \ll \lambda_D$ and $dv \ll v_{th,e}$,
seems to be adequate for correctly describing the formation of a
coherent structure much larger than $dx \times dv$. To clarify
this point, we made a number of other runs with different
numerical accuracy, corresponding to different computational
times. For the algorithms here used, we know that for the same
grid spacing the computational CPU time scales as:

\begin{equation}\label{Tau}
\tau_{_{VL3}} = \frac{3}{2} \, \tau_{_{VL2}} , \;\;\;
\tau_{_{Spline}} = 3\tau_{_{VL2}}
\end{equation}

\noindent
where $\tau_{_{VL2}}$, $\tau_{_{VL3}}$ and $\tau_{_{Spline}}$ are
the computational time for VL2, VL3 and Spline, respectively. The
criterion we used is to take a numerical accuracy for the three
algorithms corresponding to the same computational time. We made
the new runs by using all the same parameters of the previous
runs, but with $L_x =30$ and $v_{max} = 15 v_{th,e}$. The
numerical mesh are: $N_x$=300 and $N_v$=601 for VL2 (run D),
$N_x$=200 and $N_v$=401 for VL3 (run E) and $N_x=100$ and $N_v=201$ (run
F) for Spline. We now found that the energy variation is now
asymptotically constant for all methods. However, even if VL2 has
a better resolution with respect to the other two methods,
invariants decrease more than in VL3 or Spline. So, even at parity
of computational time conditions, the fact that VL3 and Spline are
third order methods while VL2 is a second order one has a relevant
effect on the invariants. Finally, VL3 and Spline have different
trends for invariants, but the same asymptotic values.

In conclusion, choosing a numerical algorithm means to select a
determined quantity of artificial dissipation. This means that,
even if the grid length scales are by far the shorter length
scales of the system, the final state of the
system can be affected by the kind of algorithm we have used.
Furthermore, even by performing very accurate simulations,
i.e. grid scales sufficiently short to have a "correct" long
time behavior of the energy and the invariants, the
long time nonlinear dynamics can be significantly different,
in particular when more "turbulent" systems are studied
(for example if some external forcing continues to
inject energy during the nonlinear regime).

This work was supported, in part, by MURST.
Laura Galeotti is pleased to acknowledge the
INFM Parallel Computing Initiative for supporting
her doctoral fellowship at Pisa University and for
giving the access to computing facilities.

\end{document}